\documentclass[lettersize,journal]{IEEEtran}
\usepackage{amsmath,amssymb,amsfonts}
\usepackage{amsthm}
\usepackage{bm}
\usepackage{algorithmic}
\usepackage{algorithm}
\usepackage{array}
\usepackage[caption=false,font=normalsize,labelfont=sf,textfont=sf]{subfig}
\usepackage{textcomp}
\usepackage{stfloats}
\usepackage{url}
\usepackage{booktabs}
\usepackage{verbatim}
\usepackage{graphicx}
\usepackage{tabularx}
\usepackage{cite}
\usepackage[colorlinks,
            linkcolor=blue,       
            anchorcolor=blue,  
            citecolor=blue,        
            ]{hyperref}
\theoremstyle{definition}

\theoremstyle{remark}

\usepackage{graphicx}
\usepackage{multirow}
\allowdisplaybreaks[4]

\begin{document}

\title{NetWorld: Communication-Based Diffusion World Model for Multi-Agent Reinforcement Learning in Wireless Networks}
\author{Kechen Meng, Rongpeng Li, Yansha Deng, Zhifeng Zhao, and Honggang Zhang
\thanks{K. Meng and R. Li are with the College of Information Science and Electronic Engineering, Zhejiang University (email: \{mengkechen, lirongpeng\}@zju.edu.cn). Y. Deng is with the Department of Engineering, King’s College London, London WC2R 2LS, U.K. (e-mail: {yansha.deng}@kcl.ac.uk). Z. Zhao is with Zhejiang Lab as well as the College of Information Science and Electronic Engineering, Zhejiang University (email: zhaozf@zhejianglab.org). H. Zhang is with the School of Computer Science and Engineering, Macau University of Science and Technology, Macau,
China (email: hgzhang@must.edu.mo).  }
}



\maketitle

\begin{abstract}
As wireless communication networks grow in scale and complexity, diverse resource allocation tasks become increasingly critical. Multi-Agent Reinforcement Learning (MARL) provides a promising solution for distributed control, yet it often requires costly real-world interactions and lacks generalization across diverse tasks. Meanwhile, recent advances in Diffusion Models (DMs) have demonstrated strong capabilities in modeling complex dynamics and supporting high-fidelity simulation. Motivated by these challenges and opportunities, we propose a Communication-based Diffusion World Model (NetWorld) to enable few-shot generalization across heterogeneous MARL tasks in wireless networks. To improve applicability to large-scale distributed networks, NetWorld adopts the Distributed Training with Decentralized Execution (DTDE) paradigm and is organized into a two-stage framework: (i) pre-training a classifier-guided conditional diffusion world model on multi-task offline datasets, and (ii) performing trajectory planning entirely within this world model to avoid additional online interaction. Cross-task heterogeneity is handled via shared latent processing for observations, two-hot discretization for task-specific actions and rewards, and an inverse dynamics model for action recovery. We further introduce a lightweight Mean Field (MF) communication mechanism to reduce non-stationarity and promote coordinated behaviors with low overhead. Experiments on three representative tasks demonstrate improved performance and sample efficiency over MARL baselines, indicating strong scalability and practical potential for wireless network optimization.
\end{abstract}


\section{Introduction}
\IEEEPARstart{W}{ireless} networks, characterized by their flexibility, low deployment cost, and robustness, have evolved into indispensable infrastructure for ubiquitous connectivity. To sustain efficient operation, numerous radio resource management tasks must be executed, such as Coordinated BeamForming (CBF) at the physical layer\cite{ge2020deep}, Resource Block (RB) scheduling at the MAC layer\cite{meng_conditional_2025a}, and Network Slicing (NS) at the network layer\cite{shao2021graph}. As network architectures become increasingly dense and heterogeneous, these tasks become even more challenging due to highly dynamic wireless channels and the inherently distributed nature of large-scale deployments, where controllers only observe partial information while their decisions remain strongly coupled. This naturally calls for coordinated and adaptive multi-agent decision-making, making Multi-Agent Reinforcement Learning (MARL) a promising paradigm for intelligent wireless resource management\cite{shao2021graph, ge2020deep, meng_conditional_2025a}. Nonetheless, most MARL methods still suffer from unstable training and low sample efficiency, due to their reliance on costly online interactions with the environment\cite{Chandra2025DiWADP}. Moreover, existing approaches are often tailored to specific tasks and generalize poorly to unseen operating conditions, since different management tasks induce heterogeneous state, action, and reward spaces with distinct dimensions and value ranges. As a result, each new deployment encounters fundamentally different dynamics in the observable space, necessitating learning from scratch, which is computationally demanding 
\cite{hafner2025mastering}. These limitations motivate the development of a \emph{shared} latent world model that reduces superficial deployment-specific differences while capturing the deeply shared commonality. 

Recent world modeling approaches increasingly adopt Recurrent State-Space Models (RSSMs)\cite{hafner2025mastering} and Transformer-based\cite{zhang2023storm} architectures to capture environment dynamics in a latent space. In most of these methods, the temporal evolution of the environment is represented as a sequence of discrete latent variables to stabilize training and facilitate planning. However, such discretization may discard fine-grained information about the underlying states, thereby limiting model expressiveness, reconstruction fidelity, and cross-task generality. 
By iteratively transforming noise into data through a learned denoising process, Diffusion Models (DMs) can approximate complex, multi-modal distributions with high sample quality and favorable convergence behavior, making them a compelling alternative for world modeling. Importantly, the induced world model contributes to reducing expensive and risky real-world interactions to obtain a readily applicable RL.

\begin{figure*}[tbp]
\setlength{\abovecaptionskip}{0cm} 
\setlength{\belowcaptionskip}{-0.5cm} 
\centering
\includegraphics[width=\textwidth]{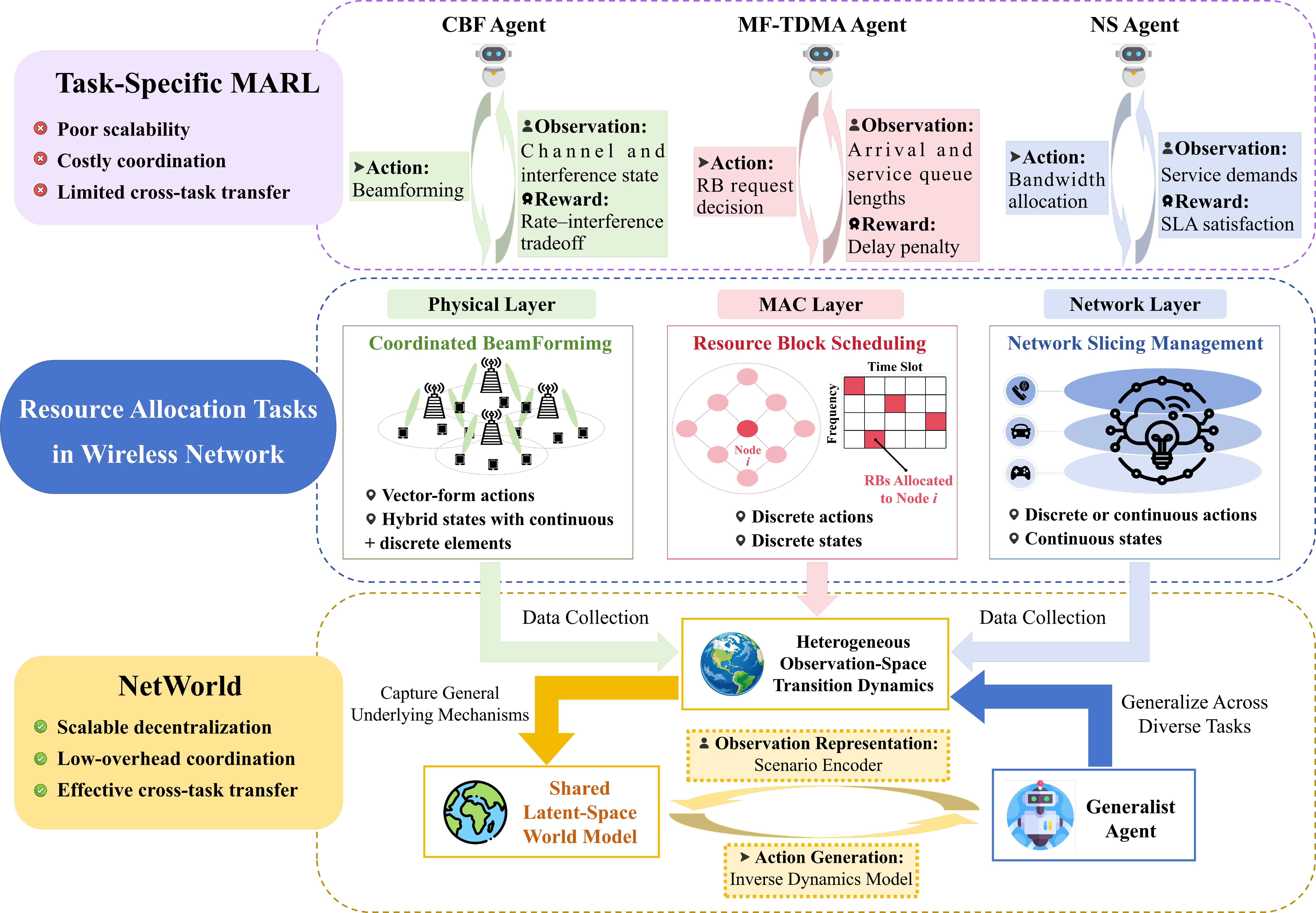}
\caption{Comparison of task-specific MARL and NetWorld paradigms in diverse wireless resource allocation tasks. }
\label{world_vs_specific}
\vspace{-0.2cm}
\end{figure*}

In this paper, contingent on \emph{multi-task offline datasets} collected from heterogeneous wireless resource allocation tasks, we envision a multi-task multi-agent diffusion world model NetWorld. By supporting few-shot adaptation, NetWorld aims to benefit the applicability of MARL in these large-scale tasks with improved data efficiency. Nevertheless, to realize such an appealing and ambitious vision, as illustrated in Fig.~\ref{world_vs_specific}, 
the development of NetWorld faces multi-folded practical challenges:
\begin{itemize}
\item \textbf{Scalable and efficient training:} The modeling and decision space grows rapidly with the agent population, and the direct concatenation of observations from all agents should be prohibited due to the curse of dimensionality.
\item \textbf{Deployable coordination:} Stringent constraints on latency and communication budgets limit information exchange, which complicates cooperation in dense networks.
\item \textbf{Cross-task heterogeneity:} The state, action, reward spaces, and even agent population differ significantly across tasks, leading to mismatched observable dynamics and making the generalization a non-trivial implementation.
\end{itemize}
Correspondingly, 
NetWorld is primarily built under the Distributed Training with Decentralized Execution (DTDE) paradigm \cite{ma2024efficient}, where each node acts as an autonomous agent with local observations and minimal communicated information. For each target task, NetWorld first learns a return-conditioned diffusion dynamics model and uses classifier guidance \cite{janner2022planning} to generate high-return trajectory segments. Afterward, it applies a task-specific inverse dynamics model to recover executable actions from the denoised transitions. To enable coordination under limited information exchange, we incorporate the Mean Field (MF) local communication mechanism \cite{yang2018mean} into the classifier, allowing guidance to depend on each agent and the averaged 1-hop neighborhood effect. Finally, NetWorld mitigates cross-task heterogeneity in observations, actions, and rewards via efficient encoding modules, enabling unified latent dynamics modeling across tasks with low deployment overhead.

\begin{figure*}[tbp]
\setlength{\abovecaptionskip}{0cm} 
\setlength{\belowcaptionskip}{-0.5cm} 
\centering
\includegraphics[width=\textwidth]{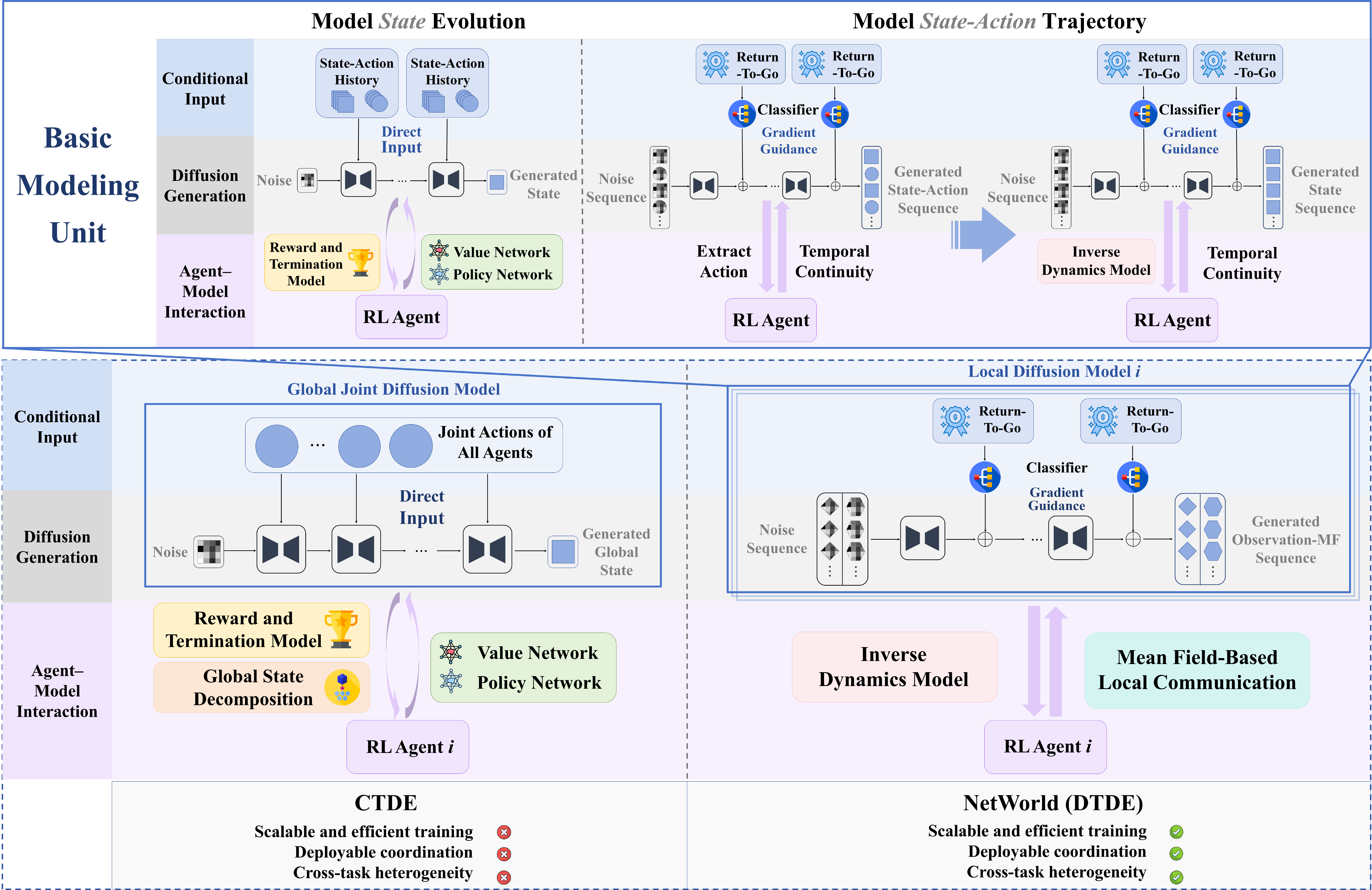}
\caption{Overview of the scalable and adaptive NetWorld framework. }
\label{single2multi}
\vspace{-0.2cm}
\end{figure*}

\section{Scalable and Adaptable Framework}
The learning of NetWorld is purely contingent on an offline dataset, where each task is formulated as a Decentralized Partially Observable Markov Decision Process (Dec-POMDP). In this context, the dynamic environment evolves with an underlying global \emph{state}, but each node makes decisions based on partially available information \cite{yang2018mean}. For each task, the learning objective is to learn a latent dynamics model from logged trajectories and enable task-specific policy adaptation to maximize the expected discounted return. 
To further accomplish the omnipotent support for distributed and large-scale multi-agent tasks, an essential prerequisite belongs to establishing a scalable and adaptive framework. 

Most recent diffusion-based approaches in multi-agent settings\cite{zhang2025revisiting} adopt the Centralized Training with Decentralized Execution (CTDE) paradigm, wherein a diffusion world model is used to capture the \emph{joint} evolution of the entire agent population, while a learned, centralized decomposer maps the global state into per-agent local observations to enable decentralized execution \cite{zhang2025revisiting}. 
Unfortunately, the boosted computational complexity \cite{ma2024efficient} in large-scale tasks and the struggling accommodation of agent population variations across tasks indicate that CTDE is not a competent candidate for NetWorld. Instead, NetWorld selects DTDE with calibrated coordination. On one hand, as its name implies, each agent in DTDE is trained and executes decisions independently. The decentralized architecture improves scalability and guarantees robustness to varying agent populations. On the other hand, the minimal communicated information avoids non-stationary dynamics in directly extending single-agent solutions \cite{alonso2024diffusion, janner2022planning}. Fig.~\ref{single2multi} highlights the methodology for the scalable and adaptive framework used in NetWorld. 

\subsection{Lightweight Dynamics Modeling}
NetWorld adopts diffusion-based environment modeling at the individual agent level, treating each agent’s local trajectory as the basic modeling unit. 
Though DMs ensure stable optimization and accurate dynamics representation via gradual \emph{forward} noising and \emph{reverse} denoising, scalable training still hinges on keeping the per-agent model compact while retaining sufficient expressive power.
Based on the modeling target of the diffusive sample, 
promising solutions can be broadly grouped into two categories, as illustrated at the top of Fig.~\ref{single2multi}.  

\begin{itemize}
\item The first category learns a transition model for \emph{state} evolution conditioned on the agent’s state–action history, and couples it with an additional \emph{reward} and \emph{termination} predictor to form a complete environment surrogate. An agent is then trained by interacting with this learned environment, typically following an actor–critic paradigm with separate value and policy function approximators optimized on model-generated rollouts \cite{alonso2024diffusion}. While effective, this line of methods introduces an extra modeling stage and relies on multi-step imagined interaction, which can accumulate model bias, amplify compounding errors over long horizons, and increase training complexity. 

\item The second category directly employs conditional diffusion to model the distribution of \emph{state–action trajectories}. Conditioning can be implemented via classifier guidance \cite{janner2022planning} or classifier-free guidance \cite{ajay2022conditional}, steering the generative process toward sequences that satisfy task constraints or yield high returns. In classifier-free guidance, a unified architecture is trained to support both conditional and unconditional denoising, and samples are guided by appropriately combining these two predictions. By contrast, classifier guidance introduces an additional classifier
, whose \emph{gradients} with respect to the current noisy sample are used to bias the denoising direction toward the desired condition, often yielding stronger controllability and sharper alignment with the conditioning objective. Nevertheless, accurately generating actions may be challenging when control decisions are discrete or change at high frequency, a typical phenomenon for tasks in wireless networks. A practical remedy is to restrict the diffusion process to modeling state trajectories only, while recovering the corresponding actions through an inverse dynamics model \cite{ajay2022conditional}. 
\end{itemize}
Finally, NetWorld adopts the latter category by employing conditional DMs to capture the dynamics of the state trajectory, where classifier guidance steers the denoising process toward anticipated returns. The resulting trajectories are then converted into executable controls via an inverse dynamics model, which infers the most likely action that induces each transition. This choice avoids complex modeling, reduces iterative agent–model coupling, and can lead to a simpler and more efficient training pipeline.

\begin{figure*}[!t]
\setlength{\abovecaptionskip}{0cm} 
\setlength{\belowcaptionskip}{-0.5cm} 
\centering
\includegraphics[width=\textwidth]{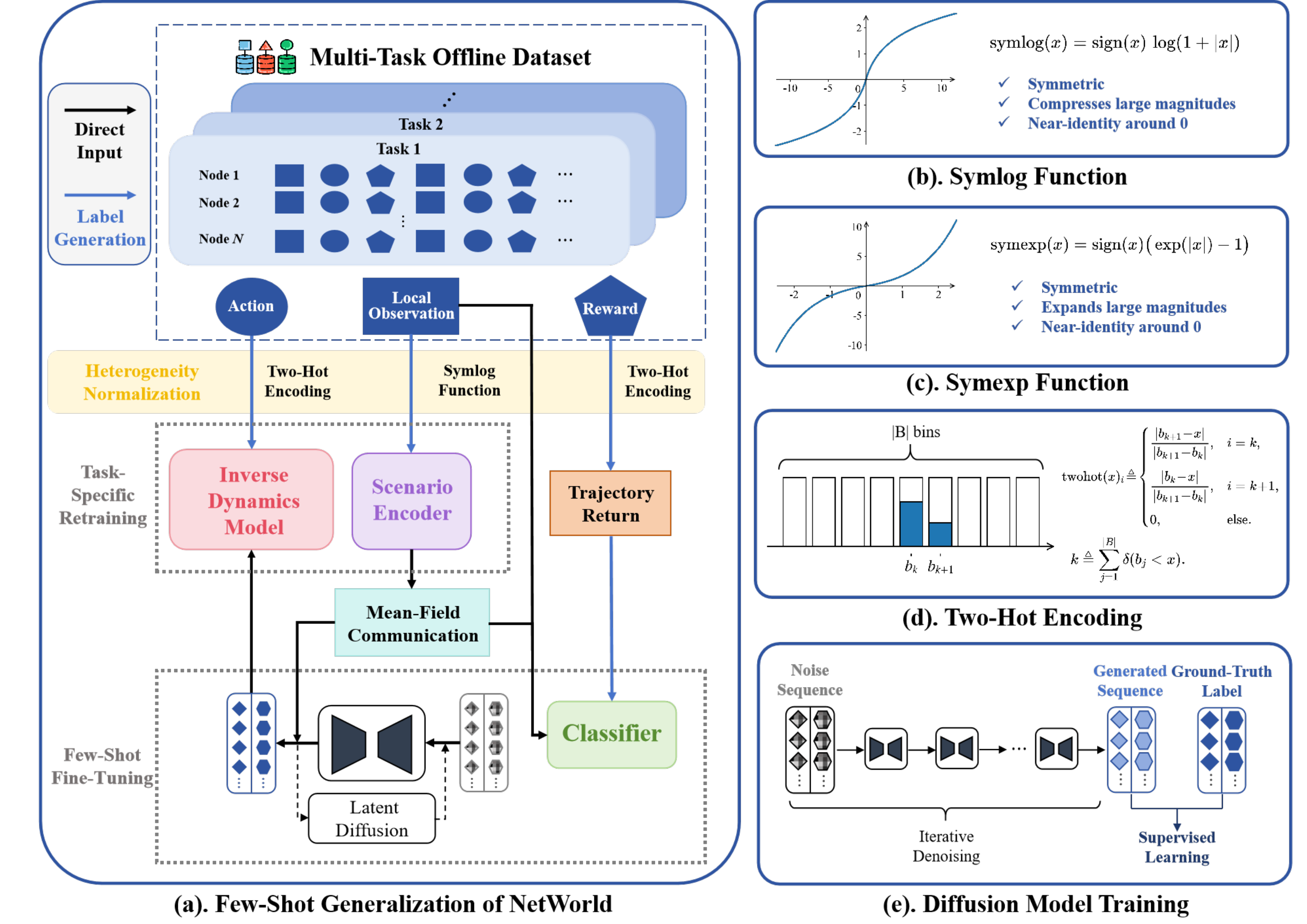}
\caption{Structural design and few-shot generalization procedure of NetWorld.}
\label{NetWorld-training}
\vspace{-0.4cm}
\end{figure*}

\subsection{Deployable Coordination}
While independent per-agent modeling improves scalability, effective coordination remains essential in multi-agent systems. Directly extending the single-agent solutions to multi-agent settings often leads to challenges such as poor scalability, susceptibility to local optima, and limited coordination. Therefore, NetWorld incorporates local communication as a complementary mechanism, allowing agents to exchange limited information without relying on global state aggregation. In practice, such local communication can be realized through several representative mechanisms. A common approach is the consensus-based method\cite{chen2022sample}, where agents propagate and aggregate information through parameterized update rules to approximate joint value functions or global coordination signals. Although convergence can be guaranteed under linear function approximation, performance often degrades when these methods are combined with deep neural networks. Another line of work employs Graph Neural Networks (GNNs) to model interactions and promote cooperation through learned relational structures\cite{nayak2023scalable}. While GNN-based approaches can capture complex dependencies, they typically suffer from high computational cost and substantial data requirements. In contrast, MF communication \cite{yang2018mean} approximates interactions in large populations by modeling those between each agent 
and the \emph{average} effect of its 1-hop neighbors,
offering both theoretical performance guarantees and improved computational scalability \cite{yang2018mean}. 
Therefore, we primarily instantiate NetWorld with MF-based communication. 

\section{NetWorld for Heterogeneous Multi-Task Wireless Control}\label{sec:NetWorld}

Different wireless resource allocation tasks, such as CBF at the physical layer\cite{ge2020deep}, RB scheduling at the MAC layer\cite{meng_conditional_2025a}, and NS at the network layer\cite{shao2021graph}, exhibit pronounced heterogeneity in their state, action, and reward spaces. Such cross-task discrepancies substantially hinder NetWorld’s transfer and few-shot adaptation. While the adoption of the DTDE paradigm with MF communication and carefully chosen modeling targets enables scalable and efficient training of a single-task world model, significant effort is still required to address cross-task mismatches. Therefore, as illustrated in Fig.~\ref{NetWorld-training}, we introduce some essential modular designs.

\begin{itemize}
\item \textbf{Task Heterogeneity-Compatible Scenario Encoding:}
To handle heterogeneous observations of different tasks, we apply the symlog function (i.e., $\mathrm{symlog}(x)=\mathrm{sign}(x)\,\log\!\left(1+\lvert x\rvert\right)$) \cite{hafner2025mastering} to each agent’s observation, which is approximately linear around the origin while logarithmically compressing large values in a sign-symmetric manner. This is crucial in our setting because the raw observation scales vary substantially across tasks. In the RB scheduling task, observations mainly describe packet generation and transmission queue lengths, so most entries are nonnegative integers with values typically below $10$. In contrast, the CBF agent observes physical-layer features such as the achievable rate, equivalent channel gains, and the total interference-plus-noise power. These features are floating-point values that are often concentrated in $[-1,1]$. In NS, observations include past and current service demands, and their magnitudes can diverge from near-zero to the order of $10^{3}$. By reducing such scale gaps, symlog makes features more comparable and stabilizes cross-task learning. 
The transformed observations are then mapped into a shared latent space using a scenario encoder. Based on this representation, the 1-hop MF approximation over neighboring agents is derived as well. The resulting latent observation sequences, together with their MF approximations, are subsequently used for diffusion modeling in a manner that is compatible with heterogeneous tasks.

\item \textbf{Task-Specific Inverse Dynamics Model:}
To generalize across tasks with different action ranges, we employ a symexp two-hot encoding loss\cite{hafner2025mastering}. Specifically, the network predicts logits over symexp-spaced bin centers (i.e., $\mathrm{symexp}(x)=\mathrm{sign}(x)\big(\exp(|x|)-1\big)$), while the target action is represented by two nonzero weights on its nearest bins. 
With symexp-spaced bins (i.e., uniform binning in the symlog domain), the discretization provides finer resolution near zero and progressively coarser resolution for large magnitudes, improving robustness to cross-task action-scale shifts.
Compared to one-hot encoding, the two-hot encoding provides smooth targets and stable gradients, which improve generalization when action ranges and distributions vary across tasks. This is particularly important in our benchmarks, where the scheduled RB action is discrete and takes nonnegative integer values typically below $10$, the CBF action is a vector-valued transmit beamformer with magnitudes around $40$, and the allocated resource for NS can span several orders of magnitude, from several kilohertz (kHz) to megahertz (MHz). We minimize a cross-entropy loss with soft targets, and reconstruct the action via a probability-weighted average over bin centers, yielding a compact discrete parameterization for continuous controls. 

\item \textbf{Cross-Task Return Conditioning:}
To overcome reward heterogeneity across tasks, we construct a unified return-based conditioning signal for classifier training. Following the two-hot discretization principle used for action modeling, we represent each scalar reward with two adjacent bins \cite{hafner2025mastering}. Such normalization effectively copes with significantly varied reward semantics and magnitudes across tasks. In RB scheduling, the reward mainly reflects delay penalties and is often negative. In CBF, it balances achievable rate gains against interference-related loss, yielding values typically around $3$. In NS, the signal is tied to the Service Level Agreement (SLA) satisfaction ratio and usually stays below $1$. For a trajectory segment of a given horizon, we then compute its discounted return and use it as the conditioning label for classifier guidance. This segment-level supervision is less sensitive to instantaneous noise, remains comparable under reward-scale shifts, and provides a more long-horizon objective to steer generation toward high-return behaviors.
\end{itemize}

\subsection{Training in NetWorld}\label{sec:training}
As summarized in Fig.~\ref{NetWorld-training}, we now describe how to train the key components that enable effective trajectory-driven planning. Since the agents within each common task are homogeneous, we employ parameter sharing across agents to improve training efficiency. 
The scenario encoder empowered by the symlog function\cite{hafner2025mastering} is trained end-to-end together with the diffusion noise model during world model learning. For consistency, the diffusion noise model 
also takes both the local latent observation sequence 
and the MF latent observation sequence 
as inputs, thereby modeling their joint distribution. The classifier 
is then trained in a supervised manner to estimate the cumulative return of a trajectory segment. In parallel, the inverse dynamics model 
is trained under a supervised objective, where the true action 
serves as the target label.  Given an offline dataset 
that aggregates trajectories from multiple tasks, we jointly optimize the classifier loss, the reverse diffusion loss, and the inverse dynamics loss. This multi-task training enables NetWorld to adapt to any new task in a few-shot manner by training the corresponding scenario encoder and inverse dynamics model only, while lightly fine-tuning the diffusion world model on a small amount of task-specific data. 

\begin{figure}[t]
\centerline{\includegraphics[width=\linewidth]{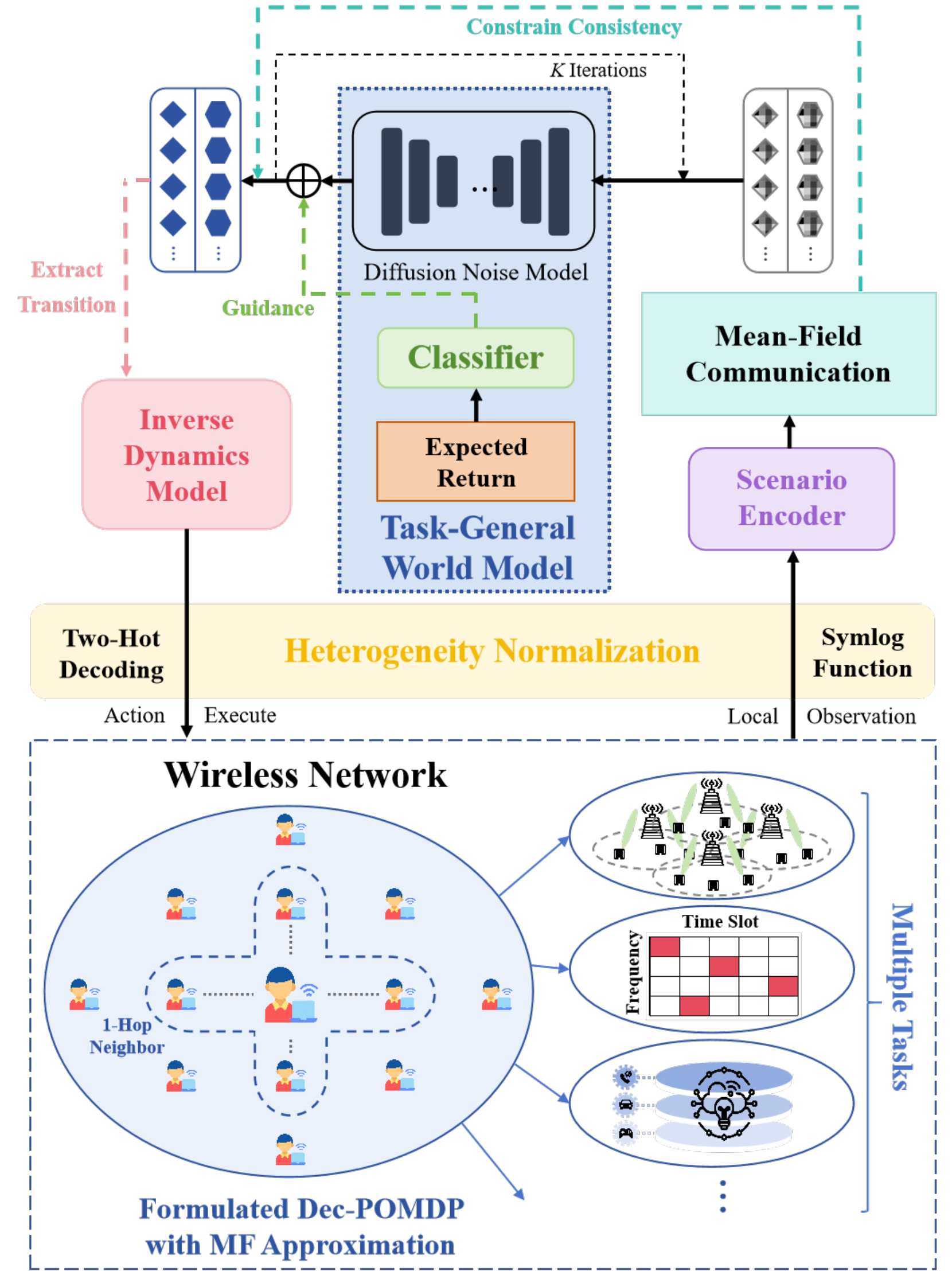}}
\vspace{-0.2cm}
\caption{Implementation procedure of NetWorld.}
\label{NetWorld-implement}
\vspace{-0.4cm}
\end{figure}

\subsection{Implementation with MF-Enhanced Classifier Guidance}\label{sec:implementation}
At the deployment stage, we fully exploit this few-shot adaptivity and employ the well-calibrated scenario encoder, classifier, diffusion noise model, and inverse dynamics model to generate trajectories for policy optimization. Specifically, each agent encodes its local observation together with the 1-hop MF approximation as the starting condition for trajectory generation. Following classifier-guided diffusion sampling \cite{janner2022planning}, the process begins from Gaussian noise and iteratively refines a specific-horizon latent trajectory that represents plausible future observation evolutions with high expected return. To ensure consistency with the real environment, the first latent observation of each trajectory is clamped to the encoded current observation at every denoising step. After some predefined denoising iterations, the model yields the next latent observation. The inverse dynamics model then maps the predicted transition to a per-agent control action, which is executed in the environment. The environment proceeds to the next state and agents receive updated observations. This procedure repeats in a receding-horizon RL control loop, enabling decentralized per-agent execution while maintaining coordinated behavior, as illustrated in Fig.~\ref{NetWorld-implement}.


\section{Feasibility Study of NetWorld}
In this section, we present numerical experiments on three typical tasks—Multi-Input Single-Output (MISO) downlink CBF \cite{ge2020deep}, Multi-Frequency Time Division Multiple Access (MF-TDMA) RB allocation \cite{meng_conditional_2025a}, and NS resource management \cite{shao2021graph} in wireless networks—to evaluate the effectiveness and verify the generalization capability of the proposed NetWorld framework.
\begin{figure}[t]
\centerline{\includegraphics[width=0.85\linewidth]{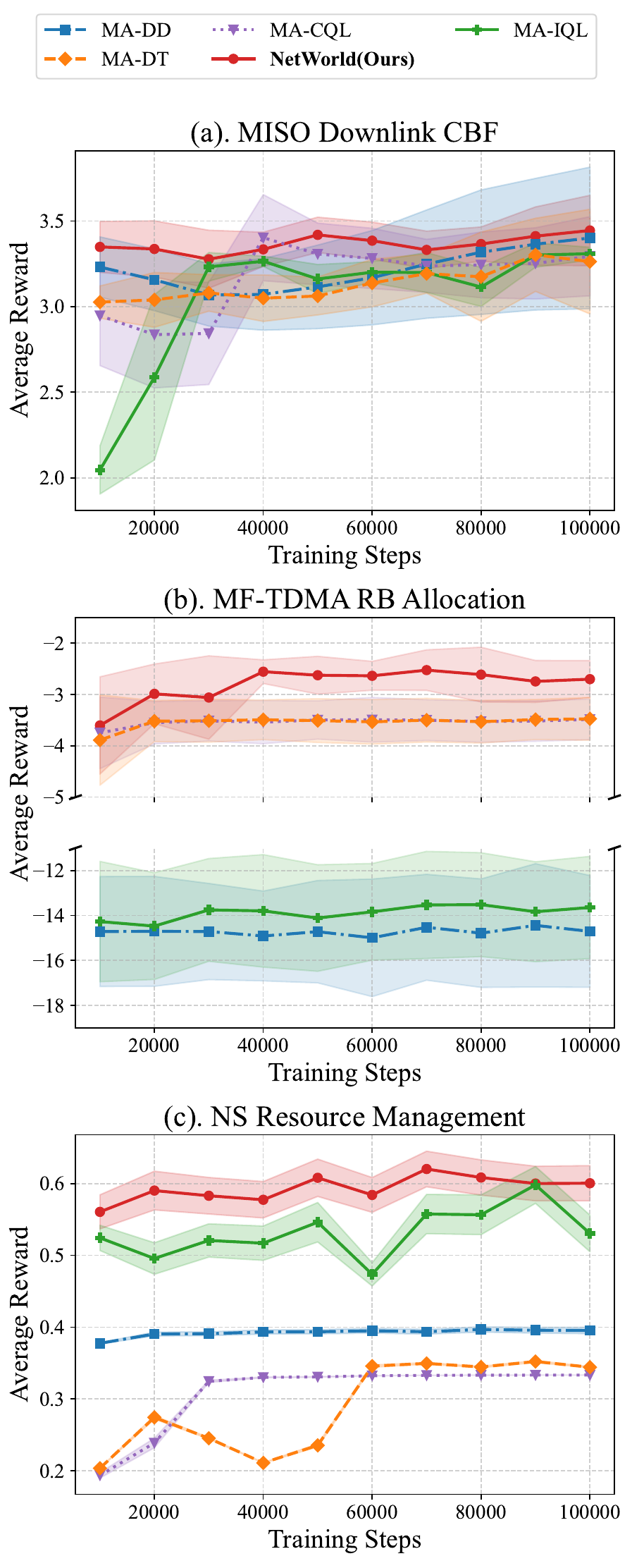}}
\vspace{-0.2cm}
\caption{Comparison of NetWorld with different methods across the three typical resource allocation tasks in terms of average reward.}
\label{superiority}
\vspace{-0.4cm}
\end{figure}
\subsection{Experimental Settings}
We implement the scenario encoder 
for local observations using a $3$-layered Multi-Layer Perceptron (MLP). The diffusion model 
follows the widely adopted Denoising Diffusion Probabilistic Model (DDPM) \cite{ho2020denoising} and is realized as a temporal U-Net\cite{janner2022planning} with $8$ residual blocks, arranged into $4$ down-sampling layers and $4$ symmetric up-sampling layers. Timestep and task index embeddings are concatenated and injected into the first temporal convolution of each block. The classifier 
adopts the down-sampling path of this U-Net and appends a final linear layer to produce a scalar output. The inverse dynamics model 
is designed as a feed-forward neural network that maps the encoded observation transitions to actions. To improve efficiency and training stability, all these networks share parameters across homogeneous agents.

We collect training data from the aforementioned MISO downlink CBF \cite{ge2020deep}, MF-TDMA RB allocation \cite{meng_conditional_2025a}, and NS resource management \cite{shao2021graph} tasks. The MF-TDMA RB allocation task operates with $8-9$ agents, whereas the other two tasks consist of $19$ agents. For each task, we first pre-train the diffusion world model on $100$ expert trajectories from the other two tasks (each trajectory contains $5,000$ steps). We then use $10$ trajectories from the held-out task to train the corresponding scenario encoder and inverse dynamics model, and to fine-tune the world model, thereby evaluating the few-shot generalization capability of the proposed framework. The baselines include several state-of-the-art offline MARL methods under the DTDE paradigm, including MA-CQL, MA-IQL, MA-DT, and MA-DD. Each baseline learns a task-specific policy from $300$ expert trajectories in the corresponding task. All methods are trained for $100$ epochs, with $1{,}000$ training steps per epoch. During evaluation, we measure performance over episodes of $1{,}000$ steps using the average reward, and report the mean and standard deviation over $3$ independent seeds to ensure statistical reliability.

\subsection{Numerical Results}

\begin{figure}[t]
\centerline{\includegraphics[width=\linewidth]{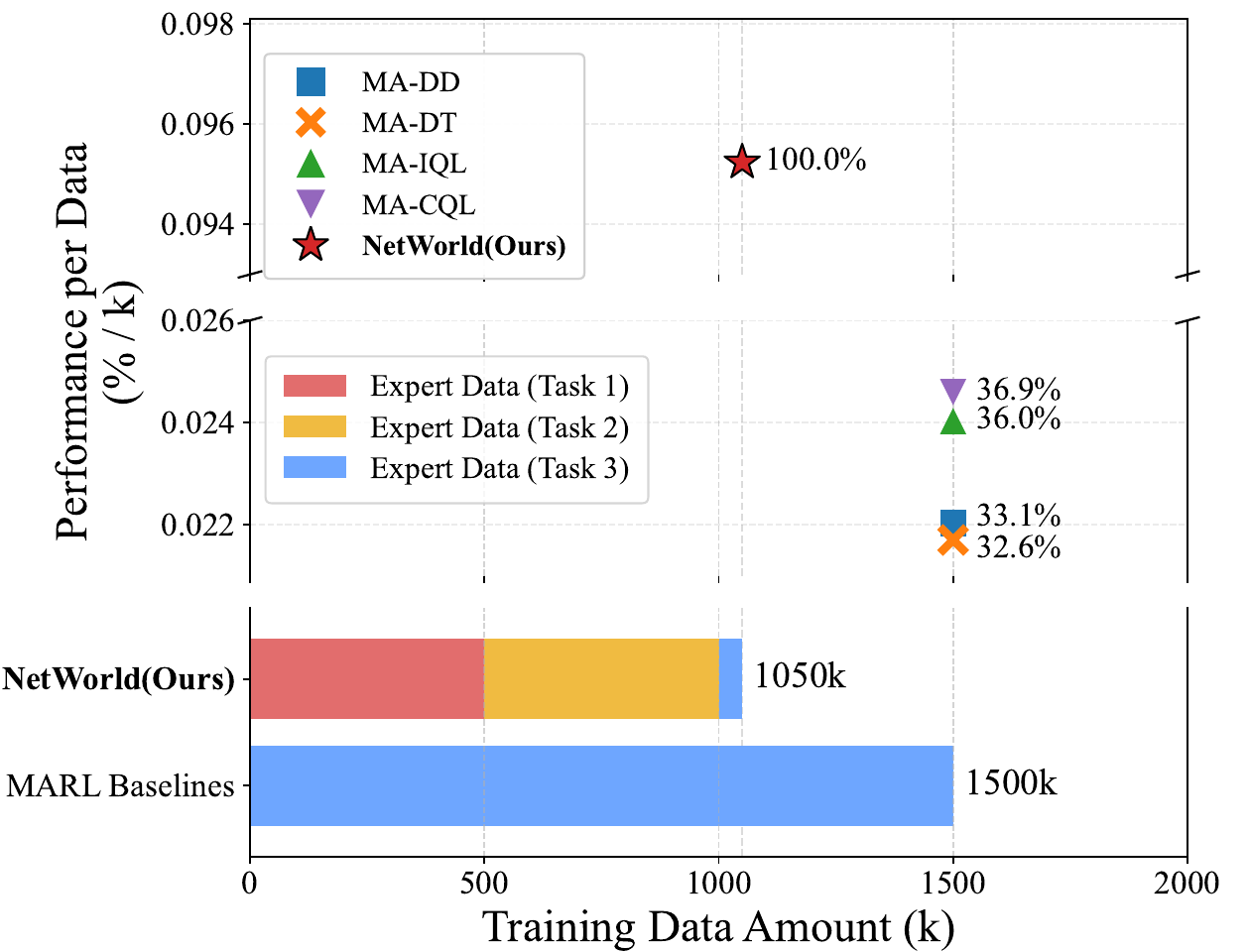}}
\vspace{-0.2cm}
\caption{Final performance–data ratio and offline dataset breakdown.}
\label{sample-efficiency}
\vspace{-0.4cm}
\end{figure}

\subsubsection{Performance Comparison}
To demonstrate the superiority of the proposed NetWorld over existing MARL methods, we conduct comparative experiments, as illustrated in Fig.~\ref{superiority}. Among the three typical resource allocation tasks, NetWorld achieves higher average rewards and exhibits smoother training curves after sufficient training. These results confirm the effectiveness of our framework design and indicate that designing the world model with conditional diffusion and MF-based local information exchange strengthens inter-agent cooperation. Moreover, the proposed approach stabilizes the learning process and generalizes well to diverse resource allocation tasks, thereby validating its broad applicability and practical effectiveness.

\subsubsection{Sample Efficiency Analysis}
We perform a few-shot learning experiment to validate the superior sample efficiency of NetWorld when adapting to a new task from the other two tasks. The performance–offline training data ratio is presented in Fig.~\ref{sample-efficiency}. 
For comparison, we report a normalized performance score computed from each method’s average reward via task-wise min–max normalization and then averaged over all tasks. The lower panel reports the offline data composition used for training. On the unseen target task, the baseline methods require approximately $1,500,000$ interaction steps to reach the reported performance level. In contrast, after pre-training the diffusion world model on expert data from the other two tasks, NetWorld achieves higher performance with roughly one-thirtieth of the expert data from the target task only. These results demonstrate that NetWorld effectively leverages knowledge captured in the learned world model to improve learning and decision-making, thereby avoiding costly and potentially unsafe data collection and exhibiting strong few-shot generalization capability.

\section{Open Research Directions}
NetWorld has provided a promising foundation for scalable and transferable MARL in wireless networks, yet several open problems remain.
\begin{itemize}
\item \textbf{Online Adaptation}: Moving from purely offline learning to deployment-aware adaptation is crucial, since real networks are non-stationary and offline data may not cover future conditions. Incorporating limited online interaction and uncertainty-aware model updates could improve robustness while controlling exploration risks. 
\item \textbf{Enhanced coordination}: 
Lightweight local exchanges scale well, but may miss fine-grained dependencies in dense and highly coupled systems. More expressive yet efficient mechanisms could enhance cooperation without violating latency or bandwidth limits. 
\item \textbf{Flexible control}: Extending diffusion-based modeling toward cross-layer and cross-timescale control may unlock broader applicability, enabling coherent decisions from fast physical-layer adaptation to slower network-level orchestration.
\end{itemize}

\section{Conclusions}
This article has presented NetWorld, a communication-based diffusion world modeling framework for heterogeneous MARL tasks in wireless networks. NetWorld enables scalable policy learning and few-shot adaptation across heterogeneous tasks, based on DTDE with return-conditioned diffusion dynamics and efficient MF-based local communications. Through multiple calibrated encoders, the framework systematically addresses inter-task heterogeneity in observations, actions, rewards, and agent populations, while remaining compatible with decentralized execution. Empirical results demonstrate that NetWorld achieves improved data efficiency and stronger generalization over conventional task-specific MARL approaches. Together, these findings highlight the potential of diffusion-based world modeling as a foundation for deployable and transferable intelligence in future wireless networks.

\end{document}